\documentclass[11pt]{article}
\usepackage{eurosym}
\usepackage{amsfonts}
\usepackage{amssymb}
\usepackage{amsmath}
\usepackage{latexsym}
\usepackage{float,subfigure,color,subfigure}
\usepackage{graphicx}
\usepackage{verbatim}
\usepackage{graphicx}
\usepackage{upref}
\usepackage{a4wide}

\newcommand{\ub}{\bar{u}}
\newcommand{\uu}{\mathbf{u}}
\newcommand{\uub}{\bar{\mathbf{u}}}
\newcommand{\Sfrac}[2]{{\textstyle{\frac{#1}{#2}}}}
\newcommand{\ddd}[1]{{\frac{\partial}{\partial #1}}}

\newcommand{\ohf}{\frac{1}{2}}

\newcommand{\tx}{\tilde{x}}
\newcommand{\ty}{\tilde{y}}
\newcommand{\tz}{\tilde{z}}
\newcommand{\ttt}{\tilde{t}}

\newcommand{\tilh}{\tilde{h}}
\newcommand{\tw}{\tilde{w}}
\newcommand{\tu}{\tilde{u}}
\newcommand{\tv}{\tilde{v}}
\newcommand{\tub}{\tilde{\ub}}
\newcommand{\tvb}{\tilde{\bar{v}}}
\newcommand{\tuu}{\tilde{\uu}}
\newcommand{\tbbu}{\tilde{\uub}}

\newcommand{\bvphi}{\boldsymbol{\varphi}}

\newcommand{\scrO}{{\mathcal{O}}}

\begin{document}

\title{A Kinematic Conservation Law in Free Surface Flow}
\author{Sergey Gavrilyuk\thanks{\texttt{sergey.gavrilyuk@univ-amu.fr}} \\
{\small Aix Marseille Universit\'{e}, CNRS UMR 7343, IUSTI, 5 rue E. Fermi} 
\\
{\small Marseille 13013, France} \\
\\
Henrik Kalisch\thanks{\texttt{henrik.kalisch@math.uib.no}}~ and Zahra
Khorsand\thanks{\texttt{zahra.khorsand@math.uib.no}} \\
{\small Department of Mathematics, University of Bergen} \\
{\small Postbox 7800, 5020 Bergen, Norway}}

\maketitle

\begin{abstract}
The Green-Naghdi system is used to model highly nonlinear weakly dispersive
waves propagating at the surface of a shallow layer of a perfect fluid. The
system has three associated conservation laws which describe the
conservation of mass, momentum, and energy due to the surface wave motion.
In addition, the system features a fourth conservation law which is the main
focus of this note. It will be shown how this fourth conservation law can be
interpreted in terms of a concrete kinematic quantity connected to 
the evolution of the tangent velocity at the free surface.
The equation for the tangent velocity is first derived 
for the full Euler equations in both two and three dimensional flows,
and in both cases, it gives rise to an approximate balance law
in the Green-Naghdi theory which turns out to be identical
to the fourth conservation law for this system.
\end{abstract}

\section{Introduction}
\label{section1}
The surface water wave problem considered here concerns the motion of an inviscid
incompressible fluid with a free surface and over an impenetrable rigid
bottom. The mathematical description of this motion involves the Euler
equations for perfect fluids coupled with free-surface boundary conditions.
Solving these equations presents a rather complex mathematical problem, and
in many instances, it can be expedient to use an approximate model for the
description of the free surface and the motion of the fluid below.
The classical simplifications of this problem are the linear theory which
may be used for waves of small amplitude, 
and the shallow water theory, which may be used for
long waves. A third class of approximations can be found in the so-called
Boussinesq scaling which requires a certain relationship between the
amplitude and wavelength of the waves to be described 

The Green-Naghdi system was derived as a long-wave model for surface water
waves which are long, but which may not necessarily have small amplitude. 
In fact, the Green-Naghdi system was first found in the special case of 
one-dimensional waves over a flat bottom \cite{Serre53}, in which case
it can be thought of as a higher-order nonlinear shallow-water system.
The system was later extended to two-dimensional surface waves \cite{Su69},
and put into the general context of fluid sheets by \cite{Green74, Green76}.
If $l$ is a characteristic wavelength and $h_0$ is the the mean water depth,
we define the dimensionless small parameter $\beta= h_0 /l$.
The Green-Naghdi equations are obtained by depth-averaging the Euler system and
truncating the resulting set of equations at second order in $\beta$;
without making any assumptions on the amplitude of the waves.

While the studies mentioned so far give a formal justification of the
Green-Naghdi system as a wave model accommodating larger amplitude waves, a
mathematical justification of this property was given for the
Green-Naghdi and some related systems in
\cite{LannesBOOK, Lannes09, YiLi2, makarenko, Saut}. Recent
years have seen increased activity in both the study of modeling
properties of the Green-Naghdi system 
(\cite{BGT,Barth04,Bonneton11,Carter11,Dias10,El06,GT,Khorsand14,YiLi1}) and in the
development of numerical discretization techniques, such as in
\cite{Cien06, Dutykh13, LGLX, Metayer10}.

The Green-Naghdi system is usually written in terms of the average horizontal
velocity $\bar{u}(t,x)$ and the total flow depth $h(t,x) = h_0 + \eta(t,x)$,
and take the form 
\begin{equation}  \label{GN_1}
h_t+\left(h\bar{u}\right)_x=0,
\end{equation}
\begin{equation}  \label{GN_2}
\bar{u}_t+ \bar{u}\bar{u}_x+gh_x- \frac{1}{3h}\frac{\partial}{\partial x}
\Big\{ h^3 \big( \bar{u}_{xt}+ \bar{u}\bar{u}_{xx}- \bar{u}_x^2 \big) \Big\}=0.
\end{equation}
Note that if higher-order terms in \eqref{GN_2} are discarded, the
system reduces to the classical shallow-water system.
On the other hand, considering waves of infinitesimally small amplitude, one
obtains the linearization of the Green-Naghdi system, which is identical to
the linearized classical Boussinesq system \cite{Per80}.

Smooth solutions of \eqref{GN_1}, \eqref{GN_2} also satisfy the
following conservation laws: 
\begin{equation}  \label{mass}
\ddd{t} \{h\} + \ddd{x} \big\{ h\bar{u} \big\} = 0,
\end{equation}%
\begin{equation}  \label{momentum}
\ddd{t} \big\{ h \bar{u}\big\} + \ddd{x} \Big\{ \frac{1}{2}gh^2 + h\bar{u}^2 - 
\frac{1}{3}h^3 \bar{u}_{xt} + \frac{1}{3} h^3 \bar{u}_x^2 - \frac{1}{3} h^3 
\bar{u} \bar{u}_{xx} \Big\} = 0,
\end{equation}%
\begin{equation}  \label{energy}
\ddd{t} \Big\{ \frac{1}{2} h \left( gh + \bar{u}^2 + {\textstyle \frac{1}{3}} 
                                        h^2 \bar{u}_x^2 \right) \Big\} 
+ \ddd{x} \Big\{ h\bar{u} \Big( gh + {\textstyle \frac{1}{2}}\bar{u}^2 
                                      + {\textstyle \frac{1}{2}} h^2 \bar{u}_x^2
                                      - {\textstyle \frac{1}{3}}h^2(\bar{u}_{xt} 
                                      + \bar{u} \bar{u}_{xx} ) \Big)  \Big\} =0,
\end{equation}%
\begin{equation}  \label{head}
\ddd{t} \Big\{ \bar{u} -hh_x \bar{u}_x - \frac{1}{3}h^2\bar{u}_{xx}\Big\} 
              + \ddd{x} \Big\{ gh + \frac{1}{2}\bar{u}^2 - h h_x \bar{u} \bar{u}_x
              - {\textstyle \frac{1}{2}} h^2 \bar{u}_x^2 
              - {\textstyle \frac{1}{3}} h^2 \bar{u} \bar{u}_{xx} \Big\} = 0.
\end{equation}%
The conservation law \eqref{mass} is easily seen to describe mass conservation.
It can also be shown that 
\eqref{momentum} represents momentum conservation and \eqref{energy}
represents the energy conservation in the Green-Naghdi approximation. The last
conservation law \eqref{head} is not easily interpreted. Some authors have
ascribed the conservation of angular momentum to this balance equation. On
the other hand, it was shown in \cite{Salmon} that it originates from
the particle relabeling symmetry for a corresponding approximate Lagrangian
of the Euler equations.

In the present work, the focus will be on giving a precise physical meaning
to the quantities appearing in the fourth conservation law. 
As will come to light, the density appearing in the
conservation law \eqref{head} is proportional to the tangential fluid velocity
at the free interface, while the flux can be interpreted as an 
approximation of the energy per unit mass of fluid. 
The conservation law derives from a corresponding balance law
which holds exactly for the surface water-wave problem for the full Euler
equations, and can be written in Lagrangian variables as 
\begin{equation}  \label{tangent}
\frac{\partial \big( \mathbf{u\cdot }\frac{\partial \mathbf{x}}{\partial s}\big)}{\partial t}
+ \frac{\partial }{\partial s}
   \Big( \frac{p}{\rho} + g h - \frac{ \left\vert \mathbf{u}\right\vert^2}{2}\Big) = 0.
\end{equation}%
Here $\mathbf{u}$ denotes the fluid velocity, 
$\mathbf{x}(s,t)$ is the fluid particle motion, $p$ is the pressure, and $\rho$ is the density. 
The variable $s$ featuring in this balance law denotes 
a parametrization of an arc lying entirely in the
fluid domain or in the free surface. In the particular case where the arc is
located in the free surface, the term $\mathbf{u} \cdot \frac{\partial 
\mathbf{x}}{\partial s}$ can be interpreted as the tangent velocity along
the free surface, multiplied by the local element of arclength. The balance
law \eqref{tangent} provides the basis for the conservation law \eqref{head}
in the Green-Naghdi approximation. This conservation law actually holds even
for the three-dimensional water-wave problem, but in this case, an
additional term enters the corresponding identity in the Green-Naghdi
approximation, so that the identity corresponding to \eqref{head} in three
dimensions is not a pure conservation law.

The disposition of the present paper is as follows. In the next section, we
describe how to obtain the balance equation \eqref{tangent} in the context
of the full Euler equations. Then, the Green-Naghdi approximation is
considered, and the associate approximate density and flux are obtained
using a method introduced in \cite{Ali12}. In Section 4, the
three-dimensional case is considered, and Section 5 contains a brief
discussion.

\section{Kinematic balance law in two-dimensional flows}
\label{section2}
The motion of a homogeneous inviscid fluid with a free surface over a flat bottom
can be described by the Euler equations with appropriate boundary conditions. 
The unknowns are the surface elevation $\eta(t,x)$, and the horizontal and vertical fluid
velocity components $u(t,x,z)$ and $w(t,x,z)$, respectively. 
The problem may be posed on a domain $\Omega_t = \left\{(x,z) | 0 < z < h_0 + \eta(t,x) \right\}$
which extends to infinity in the positive and negative $x$-direction. 
The two-dimensional Euler equations are
\begin{eqnarray}  
\frac{\partial \mathbf{u}}{\partial t}+(\mathbf{u} \cdot \nabla)\mathbf{u}
+\frac{1}{\rho}\nabla p& = \ \mathbf{g},\ & \mbox{in} \ \Omega_t, 
\label{Euler_1}
\\
\nabla \cdot \mathbf{u}& = \ 0,\ &\mbox{in} \ \Omega_t,
\label{Euler_2}
\end{eqnarray}
where $\mathbf{u} = (u,w)$ represents the velocity field, 
and $\mathbf{g} = (0,-g)$ is the gravitational acceleration. 
The free-surface boundary
conditions are given by requiring the pressure to be equal to atmospheric
pressure at the surface, i.e. $p=p_{atm}$ if surface tension effects are
neglected, and the kinematic boundary condition 
\begin{equation*}
\eta_t + u \eta_x = w
\end{equation*}
for $z=h_0 + \eta(t,x)$. For the purposes of this paper, we work with the gauge
pressure, so that we may take $p_{atm} = 0$. 

We are interested in the integral of the velocity field along a portion of
the free surface, and its evolution in time. Let $\mathcal{L}_{t}$ be a
material arc lying entirely in the free surface (cf. Figure 1). 
We define the \textit{total drift} along the arc $\mathcal{L}_{t}$ as 
\begin{equation*}
\gamma =\int\limits_{\mathcal{L}_{t}}\mathbf{u\cdot }d\mathbf{x}.
\end{equation*}
It is convenient with a view towards the three-dimensional problem to take a
somewhat more general approach, and consider the integral of the velocity
over a material arc in the fluid. To this end, we introduce semi-Lagrangian
coordinates. Following Zakharov \cite{Zakharov}, 
the change of variables $(x,z)\rightarrow \left( x,\lambda \right) $ is
given by the function 
\begin{equation*}
z=\Phi \left( t,x,\lambda \right) ,
\end{equation*}%
where $\Phi \left( t,x,\lambda \right) $ is the solution of the equation%
\begin{equation*}
\frac{\partial \Phi }{\partial t}+\frac{\partial \Phi }{\partial x}u=w,
\end{equation*}%
with the initial condition 
\begin{equation*}
\Phi \big\vert _{t=0}=\lambda h_{0} + \lambda \eta_0(x),
\end{equation*}%
where $\eta_0(x) $ is the initial disturbance of the free surface. At each
time $t$, this change of variables represents a foliation of the domain in $%
\left( x,z\right) $ variables by contact lines indexed by $0\leq \lambda
\leq 1$, and where each line corresponds to $\lambda =const$ in the 
$(x,\lambda)$-plane. The free surface $z=h_0 + \eta( t,x) $
corresponds to the value $\lambda =1$. 
It can be shown that in semi-Lagrangian coordinates, the material derivative
is defined only in terms of the horizontal velocity, and the curvilinear
domain occupied by the fluid in the plane $\left\{ (x,z) , 0<z<h_0 +
\eta(t,x) \right\} $ maps to a strip in the plane $\left\{(x,\lambda),
0<\lambda <1 \right\}$. 
Let $\mathcal{L}_{t}^{\lambda }$ be a
finite material arc lying entirely in one of the contact lines defined
above, starting at a point 
$\mathbf{x}_A(t) = (x_A(t),z_A(t))^T$, and ending at another point 
$\mathbf{x}_B(t) = (x_B(t),z_B(t))^T$. 
The arc corresponding to the value $\lambda =1$ lies entirely in the free surface,
and will be denoted simply by $\mathcal{L}_t$ as already introduced above.
\begin{figure}[!b]
\begin{center}
\includegraphics[width=0.5\textwidth]{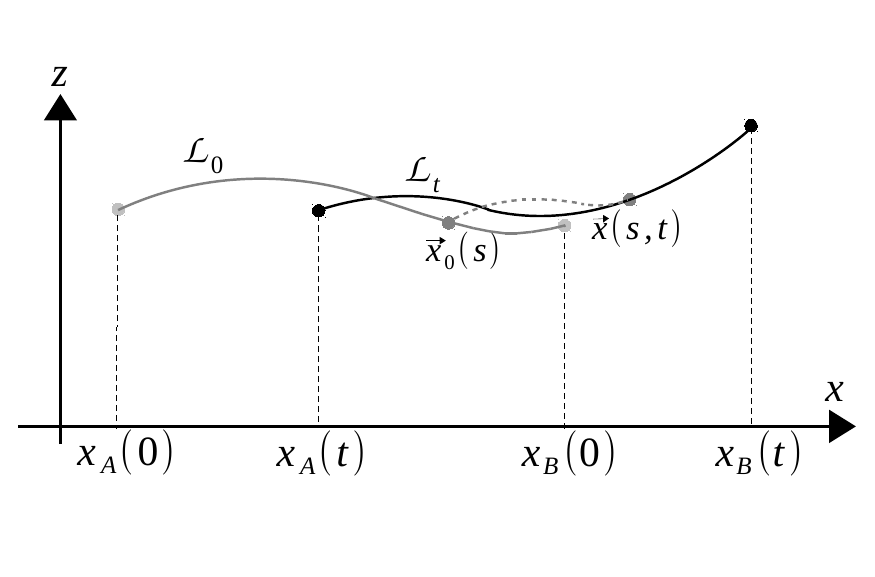}
\end{center}
\caption{This figure indicates the time evolution of an arc $\mathcal{L}_0$ which lies 
entirely in the free surface.}
\end{figure}

Parametrizing the initial arc $\mathcal{L}_{0}^{\lambda }$ using the real
parameter $s$, we obtain the description of the arc 
$\mathcal{L}_{t}^{\lambda }$ in the form 
\begin{equation*}
\mathbf{x} = \bvphi(t,s,\lambda) 
= \left( \varphi^x\left(t,s,\lambda \right),\varphi^z\left( t,s,\lambda \right) \right)^T 
,\quad s_{A}\leq s\leq s_{B},
\end{equation*}
where $\mathbf{x} =\bvphi (t,s,\lambda)$ is the solution of the Cauchy problem 
\begin{equation*}
\frac{d\mathbf{x}}{dt}=\mathbf{u}(t,\mathbf{x}),\qquad \left. 
\mathbf{x}\right\vert _{t=0}=\mathbf{x}_0(s,\lambda) ,\qquad 
\mathbf{x}_0(s,\lambda) \in \mathcal{L}_0^{\lambda}.
\end{equation*}%
Using the Euler equations \eqref{Euler_1}, \eqref{Euler_2} one can obtain the
time evolution of the total drift along $\mathcal{L}_t^{\lambda }$ in the
form 
\begin{equation}  \label{integral_relation}
\frac{d\gamma^{\lambda }}{dt} = \frac{d}{dt} \int\limits_{\mathcal{L}_t^{\lambda}}
\mathbf{u}\cdot d\mathbf{x} = \left. \Big(\frac{\left\vert 
\mathbf{u} \right\vert^2}{2}- \frac{p}{\rho} - gz\Big)\right\vert_{\mathbf{x}_A(t)}^{ \mathbf{x}_B(t)}.
\end{equation}%
One may use standard transformations to see that the integral relation
\eqref{integral_relation} can also be written as the local equation %
\eqref{tangent} along the material curve $\mathcal{L}_t^{\lambda }$.
In fact, the previous computation is nothing but the standard proof
of Kelvin's circulation theorem, but applied on a contour which is
not necessarily closed.
Taking a contour embedded in the free surface, the pressure vanishes, and we
obtain the simpler conservation law 
\begin{equation}  \label{conservation_law}
\frac{\partial 
\big( \mathbf{u} \cdot \frac{\partial \bvphi}{\partial s}\big)}{\partial t} 
  + \frac{\partial}{\partial s} 
     \bigg(g h - \frac{\left\vert \mathbf{u}\right\vert^2}{2} \bigg) = 0.
\end{equation}
In order to understand the density in \eqref{conservation_law}, we let $%
\mathbf{n=}\left( -\frac{\partial h}{\partial x},1\right) ^{T}$ be the unit
normal to the free surface, scaled by the element of arclength of the free
surface. By the same token, 
${\boldsymbol{\tau }} = \big( 1,\frac{\partial h}{ \partial x} \big)^T$ 
is the unit tangent vector to the free surface,
scaled by the element of arclength. We set 
\begin{equation*}
K=\left. \uu \cdot {\boldsymbol{\tau}}\right\vert _{z=h(t,x) },
\end{equation*}%
and observe that we have 
\begin{eqnarray*}
\mathbf{u}(t,s) \mathbf{\cdot }\frac{\partial {\boldsymbol{\varphi}}(t,s,1) }{\partial s} 
= K(t,x)\frac{ \partial \varphi^x(t,s,1)}{\partial s}
\end{eqnarray*}%
at the free surface. Finally, in Eulerian coordinates, we have 
\begin{equation}  
\label{newK}
\frac{\partial K}{\partial t}+\frac{\partial }{\partial x}\left( u K+gh-%
\frac{1}{2}(u^2 + w^2)\right) = 0.
\end{equation}%
This is an exact conservation law representing the evolution of the tangent
velocity $K$ along the free surface. 
In fact, it can be shown that $K$ falls under the scope of the work of
Benjamin and Olver \cite{BO}
if the density $T_4$ in \cite{BO} is differentiated with respect to $x$. 
However, to the best of our knowledge, the physical interpretation
of $K$ is new.

\subsection{2D balance law in the Green-Naghdi approximation}
In order to understand the approximation of \eqref{newK} in the Green-Naghdi
theory, we recall the scaling which is used in the derivation of the system
\eqref{GN_1}, \eqref{GN_2}.
Define the non-dimensional variables
\begin{equation}
\label{scaling}
\tilde{x}=\frac{x}{l},\quad \tilde{z}=\frac{z}{h_0},\quad \tilde{t}=\frac{c_0}{l}t,
\quad \tilh=\frac{h}{h_0},
\quad
\tilde{u}=\frac{u}{c_0},\quad \tilde{w}=\frac{w}{\beta c_0},
\quad
\tilde{P}=\frac{P}{\rho g h_0}.
\end{equation}
%
Using this scaling,
the quantity  $K = u + w\frac{\partial h}{\partial x}$
is given in non-dimensional variables as 
\begin{equation*}
\tilde{K} \left(\ttt,\tx\right) = \tu+\beta \tw \tilh_{\tx},
\end{equation*}
and the conservation law \eqref{newK} turns into 
\begin{equation*}
\frac{\partial\tilde{K} }{\partial \ttt}
+ \frac{\partial}{\partial \tx}
 \Big(\tilde{K} \tu + \tilh - \frac{1}{2}(\tu^2 + \beta\tw^2)\Big) = 0.
\end{equation*}
Substituting the expressions for $\tilde{u}$ and $\tilde{w}$ obtained in the
derivation of the Green-Naghdi system yields 
\begin{equation*}
\frac{\partial}{\partial \ttt}
\Big(\tub - \frac{\beta}{3}\tilh^2 \tub_{\tx\tx} 
- \beta\tilh\tilh_{\tx} \tub_{\tx} \Big)
+ \frac{\partial}{\partial \tx}
\left( \frac{1}{2} \tub^{2} - \frac{\beta}{3} 
        \tilh^2 \tub \tub_{\tx\tx} 
- \beta \tilh\tilh_{\tx}\tub \tub_{\tx} 
+ \tilh - \frac{\beta}{2} \tilh^2 \tub^2_{\tx} \right)
= \scrO(\beta^{2}).
\end{equation*}%
Using similar reasoning as in \cite{Ali12,Ali14},
we define the non-dimensional quantities $\tilde{\mathcal{K}}$ and
$\tilde{q}_{\mathcal{K}}$ by 
\begin{equation*}
\tilde{\mathcal{K}} = \Big(\tub - \frac{\beta}{3} \tilde{h}^2 \tub_{\tx\tx}
 -\beta\tilh \tilh_{\tx} \tub_{\tx}\Big)
\end{equation*}%
and 
\begin{equation*}
\tilde{q}_{\mathcal{K}}
= \left( \big( \frac{1}{2}\tub^2
        - \Sfrac{\beta}{3} \tilh^2 \tub\tub_{\tx\tx}
        - \beta \tilh\tilh_{\tx} \tub \tub_{\tx} \big) 
+ \tilh -\frac{\beta}{2}\tilh^2 \tub^2_{\tx} \right).
\end{equation*}%
Then the balance is 
\begin{equation*}
\frac{\partial \tilde{\mathcal{K}} }{\partial \tilde{t}} + \frac{\partial 
\tilde{q}_{\mathcal{K}}}{\partial \tilde{x}}=\mathcal{O}(\beta^2).
\end{equation*}%
Using the natural scalings $\mathcal{K}=c_{0}\tilde{\mathcal{K}}$ and 
$q_{\mathcal{K}} = c_{0}^{2}\tilde{q}_{\mathcal{K}}$, the dimensional forms of
these quantities are given by 
\begin{equation*}
\mathcal{K}= \bar{u}-\frac{1}{3}h^{2}\bar{u}_{xx}-hh_{x}\bar{u}_{x},
\end{equation*}
\begin{equation*}
q_{\mathcal{K}} = gh + \frac{1}{2} \bar{u}^{2}
                 -\frac{1}{3}h^{2}\bar{u}\bar{u}_{xx}
                 - hh_{x}\bar{u}\bar{u}_{x}
                 - \frac{1}{2}h^{2}\bar{u}_{x}^{2},
\end{equation*}
respectively. While the preceding analysis shows that the conservation law
\begin{equation}
\frac{\partial \mathcal{K} }{\partial t} + \frac{\partial q_{\mathcal{K}}}{\partial x}= 0,
\label{approxCirc}
\end{equation}
holds approximately to second order in $\beta$, it can be seen immediately
that the density $\mathcal{K}$ and the flux $q_{\mathcal{K}}$ are the same
as the respective quantities in \eqref{head}, 
so that \eqref{approxCirc} is an exact identity.

It is interesting to see what happens in the
shallow-water approximation where the balance 
\begin{equation*}
\frac{\partial \tilde{\mathcal{K}} }{\partial \tilde{t}} + \frac{\partial 
\tilde{q}_{\mathcal{K}}}{\partial \tilde{x}}=\mathcal{O}(\beta)
\end{equation*}
only holds to the order of $\beta$. In this case, we obtain 
\begin{equation*}
\mathcal{K}= \bar{u}
\end{equation*}
for the density and 
\begin{equation*}
q_{\mathcal{K}}= gh+\frac{1}{2} \bar{u}^{2}
\end{equation*}
for the flux. Thus in this case the equation 
$\frac{\partial \mathcal{K} }{\partial t} + \frac{\partial q_{\mathcal{K}}}{\partial x}= 0$ 
also holds exactly, and in fact it
is just the second equation 
in the shallow-water system.
It can be seen
clearly from this form that this is a balance law relating horizontal
velocity energy density per unit mass.
In the shallow-water theory, the horizontal velocity is taken to be uniform
throughout the fluid column, so that the simple form for $\mathcal{K}$ and $q_{\mathcal{K}}$ is achieved.
However, in the present theory, the horizontal velocity is assumed to vary quadratically
with the depth. This results in a more complicated set of equations, and may also be
used advantageously for kinematic studies of the fluid motion, such as in
\cite{BK,Borluk12,Khorsand14}.

%
%
%
%
%
\section{Kinematic balance law in three dimensions}
\label{section3}
The Euler equation \eqref{Euler_1}, \eqref{Euler_2} are written in general
form, and may also be interpreted to hold in three dimensions, for a velocity
field $\uu = (u,v,w)^T$, a pressure $p(t,x,y,z)$ and body forcing $\mathbf{g} = (0,0,-g)$.
In this case, the kinematic free surface boundary condition is given by
$\eta_t + u \eta_x + v \eta_y = w$ at $z = \eta(t,x,y)$.
Note that for the purpose of this last section, we define the horizontal spatial coordinates,
velocity vector field,
gradient and material derivative by
\begin{equation*}
\mathbf{x}_H =(x,y)^T, \quad \uu_H =(u,v)^T, 
\quad \nabla_H = \Big(\frac{\partial }{\partial x},\frac{\partial }{\partial y}\Big)^T,
\ \ \mbox{and} \ \ 
\frac{D_H}{Dt} = \frac{\partial }{\partial t} + \uu_H \cdot \nabla_{H},
\end{equation*}%
respectively.

An analogous vector law for the evolution of the tangent velocity can also
be derived in three-dimensional case. 
Let $\Sigma _t$ be defined by $z=h(t,\mathbf{x}_H),$  
be the free surface, which is assumed can be projected
onto the horizontal plane $(Oxy) $. As before, we will introduce
a special Lagrangian coordinate $0\leq \lambda \leq 1$ corresponding to the
foliation of the fluid domain by the material surfaces $\Sigma_t^{\lambda}$. 
The free surface $\Sigma_t$ will correspond to the value $\lambda = 1$. 
Such a foliation can be achieved by solving the Cauchy problem for the
equation 
\begin{equation*}
\frac{\partial \Phi }{\partial t}+\uu_H\cdot \nabla_H\Phi =w,
\end{equation*}
with the initial condition 
\begin{equation*}
\left. \Phi \right\vert _{t=0}=\lambda h_{0} + \lambda \eta_0(\mathbf{x}_H) ,
\end{equation*}%
and making then the change of variables $\left( \mathbf{x}_{H},z\right)
\rightarrow ( \mathbf{x}_H,\lambda) $,
$z=\Phi(t,\mathbf{x}_H,\lambda).$

Then we introduce the Lagrangian coordinates $s^{1},s^{2}$ which may be
taken as the $(x,y)$ coordinates of a particle at time $t=0.$ Each surface $%
\Sigma _{t}^{\lambda }$ is thus parametrized by the same parameters $%
s^1,s^2.$ The change of variables is assumed to be invertible: 
$s^i=s^i(t,\mathbf{x}_2,\lambda)$, for $i=1,2.$ The
Lagrangian coordinates $s^{i}$ are conserved along the particle trajectories
in the Eulerian coordinates. This implies that the transport equation for
the Lagrangian coordinates $s^i$ %
reduces to 
\begin{equation}
\Big( \frac{\partial }{\partial t}+\uu_H\cdot \nabla _{H}\Big) s^i=0,
\qquad i=1,2,  \label{conservation_Lagrangian_coordinates}
\end{equation}%
in the coordinates $(\mathbf{x}_H,\lambda )$. Let 
\begin{equation*}
\mathbf{e}^i=\nabla _Hs^i.
\end{equation*}
Taking the horizontal gradient $\nabla _H$ of
\eqref{conservation_Lagrangian_coordinates} yields 
\begin{equation}
\mathbf{e}_t^i+\nabla_H( \mathbf{e}^i \cdot \uu_H) = 0.  
\label{AdvectionEquation}
\end{equation}%
Since the matrix $\frac{\partial \mathbf{e}^i}{\partial \mathbf{x}_H}$ 
is symmetric, \eqref{AdvectionEquation} can be written in the form 
\begin{equation}
\frac{D_{H}}{Dt}\mathbf{e}_{t}^{i}=\mathbf{e}_{t}^{i}+\left( \frac{\partial 
\mathbf{e}^{i}}{\partial \mathbf{x}_{H}}\right) \mathbf{u}_{H}=-\left( \frac{%
\partial \mathbf{u}_{H}}{\partial \mathbf{x}_{H}}\right) ^{T}\mathbf{e}^{i}.
\label{non-conservative-equation}
\end{equation}%
In analogy with the two-dimensional case, one may take the drift along each $%
s^{i}-$ curve contained in the contact surface $\Sigma _{t}^{\lambda }$ to
obtain conservation laws analogous to \eqref{tangent} in the form 
\begin{equation}
\frac{\partial \big( \mathbf{u}\cdot \frac{\partial \bvphi}{\partial s^{i}}\big) }{\partial t}
 + \frac{\partial }{\partial s^{i}}\bigg( \frac{p}{\rho}
  + g z - \frac{\left\vert \mathbf{u}\right\vert ^{2}}{2}\bigg) = 0.
\label{3D circulation}
\end{equation}%
where $\bvphi(t,s^1,s^2,\lambda)$ is the motion
of particles along the surfaces $\Sigma _{t}^{\lambda }.$ Evaluating at the
free surface $\lambda =1$ yields a formula analogous to %
\eqref{conservation_law}. Defining 
$K_i = \uu \cdot \frac{\partial \bvphi}{\partial s^i}$ 
in analogy with the two-dimensional case, 
we can define the tangent component of the velocity
vector at the free surface as 
\begin{equation*}
\mathbf{K} = K_1\mathbf{e}^1 + K_2 \mathbf{e}^2.
\end{equation*}%
Taking the partial derivative of $\mathbf{K}(t,s^1,s^2) $
with respect to time at fixed Lagrangian coordinates $s^{1},s^{2}$ we obtain 
\begin{equation*}
\frac{\partial \mathbf{K}\left( t,s^{1},s^{2}\right) }{\partial t}
 = \frac{\partial K_1}{\partial t}\mathbf{e}^1 
 + \frac{\partial K_{2}}{\partial t} \mathbf{e}^2 
 + K_1 \frac{\partial \mathbf{e}^1 (t,s^1,s^2)}{\partial t}
 + K_2 \frac{\partial \mathbf{e}^2 (t,s^1,s^2)}{\partial t}
\end{equation*}%
\begin{equation*}
= - \frac{\partial }{\partial s^1} \bigg( gh - \frac{\vert \mathbf{u}\vert^2}{2}\bigg) \mathbf{e}^1
                                            -\frac{\partial }{\partial s^2}
\bigg( gh - \frac{\vert \uub \vert^2}{2} \bigg) 
\mathbf{e}^2 + K_1 \frac{\partial \mathbf{e}^1(t,s^1,s^2)}{\partial t}
             + K_2\frac{\partial \mathbf{e}^2(t,s^1,s^2)}{\partial t}.
\end{equation*}%
Converting to Eulerian coordinates, we obtain 
\begin{equation*}
\frac{D_{H}}{Dt}\mathbf{K}(t,\mathbf{x}_H) 
= - \nabla _H\Big( gh - \frac{\vert \mathbf{u}\vert^2}{2}\Big) 
  +  K_1\frac{D_H}{Dt} \mathbf{e}^1 + K_2\frac{D_H}{Dt}\mathbf{e}^2.
\end{equation*}%
Using \eqref{non-conservative-equation}, we may write the last relation as 
\begin{equation*}
\frac{D_H}{Dt} \mathbf{K} (t,\mathbf{x}_H) = 
- \nabla_H \bigg( gh - \frac{\vert \uub \vert^2}{2} \bigg) 
- \left( \frac{\partial \mathbf{u}_H}{\partial \mathbf{x}_H}\right)^T \mathbf{K}.
\end{equation*}%
This equation may also be written as
\begin{equation*}
\frac{\partial \mathbf{K}(t,\mathbf{x}_H) }{\partial t}
 + \bigg[ \frac{\partial \mathbf{K}}{\partial \mathbf{x}_H}
          -\left( \frac{\partial \mathbf{K}}{\partial \mathbf{x}_H}\right)^T \bigg] \mathbf{u}_H
 +\left( \frac{\partial \mathbf{K}}{\partial \mathbf{x}_H}\right)^T  \! 
     \mathbf{u}_H
 + \left( \frac{\partial \mathbf{u}_{H}}{\partial \mathbf{x}_{H}}\right)^T \mathbf{K}
 + \nabla_H \bigg( gh-\frac{\vert \mathbf{u} \vert^2}{2} \bigg) = 0,
\end{equation*}%
or equivalently as
\begin{equation}
\frac{\partial \mathbf{K}(t,\mathbf{x}_H) }{\partial t}
 + {\rm{curl}}(\mathbf{K})\times \mathbf{u}_H
 + \nabla_H \bigg\{ \mathbf{K\cdot u}_H
 + gh -\frac{\vert \mathbf{u} \vert^2}{2}\bigg\} = 0.
\label{equation_new}
\end{equation}%
This equation is the equivalent of \eqref{newK} in the three-dimensional
situation. 
Moreover, 
it can be shown that $\mathbf{K}$ can be written in the form
\begin{equation}
\mathbf{K}\left( t,\mathbf{x}\right) =\mathbf{u}_{H}+w\nabla _{H}h,
\label{KK}
\end{equation}%
where $\mathbf{u}_{H}$ is evaluated at the surface and $w$ is the vertical
velocity evaluated at the interface. This is a natural generalization of the
one-dimensional case considered before. 
Since the vertical velocity $w$ at the free
surface is given by 
$w=\frac{D_{H}}{Dt}\eta,$ 
\eqref{equation_new} may be rewritten in the form 
\begin{equation}
\frac{\partial \mathbf{K}}{\partial t}
+ {\rm{curl}}(\mathbf{K})\times \mathbf{u}_{H}
+ \nabla _{H}\left\{ \mathbf{K\cdot u}_{H}
+ gh - \frac{1}{2}\bigg(\left\vert \mathbf{u}_{H}\right\vert ^{2}
     +\left( {\Sfrac{D_{H}h}{Dt}}\right)^{2}\bigg) \right\} = 0.  
\label{modified_K}
\end{equation}
The equation \eqref{equation_new} or \eqref{modified_K} 
is an \textit{exact consequence of the Euler equations} 
expressed in terms of variables evaluated at the free surface. 
Since $\mathbf{K}$ is a vector in the $(x,y)-$plane (see \eqref{KK}), 
the vector $\rm{curl}(\mathbf{K})$ has only one non-zero component and 
is orthogonal to the $(x,y)-$plane.

Note that in the present three-dimensional case, the density $\mathbf{K}$
is not covered by the results of \cite{BO} 
because \eqref{modified_K} becomes a conservation law 
only if $\mathrm{curl}(\mathbf{K}) \times u_H$ = 0.

%
%
\subsection{3D balance law in the Green-Naghdi approximation}
%
%
Next, we consider the approximation of \eqref{equation_new} in the
Green-Naghdi regime. Recall that the Green-Naghdi equations for
three-dimensional flows are written in terms of
the depth-averaged horizontal velocity
$\uub$ and the total flow depth $h$ as
\begin{equation}
\label{GN_3d_1_dim}
\eta_t + \nabla_H \cdot( h \uub ) = 0,
\end{equation}
\begin{equation}
\label{GN_3d_2_dim}
\big\{1 + \mathcal{T}(\uub,h)\} _t 
+ \uub \cdot \nabla_H \uub + \nabla_H \eta + 
Q(h, \uub) = 0,
\end{equation}
where 
\begin{equation*}
\mathcal{T}(h,\uub) = -\frac{1}{3h} \nabla_H(h^3 \nabla_H \cdot \uub),
\end{equation*}
\begin{equation*}
Q(h,\uub) = \frac{-1}{3h} \nabla_H 
\Big[ h^3 \Big( \uub  \cdot \nabla_H (\nabla_H \cdot \uub) 
                -  (\nabla_H \cdot \uub )^2 \Big)\Big].
\end{equation*}
In order to obtain an unbiased representation,
we write the non-dimensionalized version of \eqref{equation_new}
in terms of the principal unknown variables of the Green-Naghdi equation.
Thus in non-dimensional form, \eqref{equation_new} becomes 
\begin{multline}
\ddd{\ttt}\big( \tuu_H + \beta \tw \tilde{\nabla}_H \tilh \big)
+ \widetilde{\mathrm{curl}} \big( \tuu_H + \beta \tw \tilde{\nabla}_H \tilh \big) \times \tuu_H
\\
+ \tilde{\nabla}_H \left\{ \tuu_H + \beta ( \tw \tilde{\nabla}_H \tilh)\cdot \tuu_H
                                + \tilh - \ohf \Big( |\tuu_H|^2 + \beta \tw^2 \Big) \right\} = 0.
\label{equation_nonD}
\end{multline}%
As shown in the appendix, the non-dimensional horizontal velocity 
field $\tuu_{H}$, and the vertical velocity component
$\tw$ are given in the Green-Naghdi approximation at any location 
$(\tilde{\mathbf{x}}_{H},\tz)$ in the fluid by the following expressions 
in terms of the depth averaged velocity:
\begin{equation*}
\tuu_{H}(\tilde{\mathbf{x}},\tz,\ttt) 
 = \tbbu + \beta \Big( \frac{\tilh^2}{6} - \frac{\tz^2}{2} \Big) 
 \tilde{\Delta}_{H}\tbbu + \scrO(\beta^2),
\end{equation*}%
\begin{equation*}
\tw(\tilde{\mathbf{x}}_{H},\tz,\ttt) 
= -\tz\left( \nabla _{H}\cdot \tbbu \right) + \scrO(\beta).
\end{equation*}%
Moreover, in the Appendix, the following important relation is established:
\begin{equation*}
\tilde{\Delta}_H\tilde{\bar{\mathbf{u}}}
= \tilde{\nabla}_H ( \tilde{\nabla}_H \cdot \tbbu ) + \scrO(\beta).
\end{equation*}%
This relation holds as long as irrotational solutions of Euler's
equations are considered.
We now look at the three terms in equation \eqref{equation_nonD} individually.
The first term can be written as
\begin{equation*}
\ddd{\ttt}\big( \tuu_H + \beta \tw \tilde{\nabla}_H \tilh \big)
= \ddd{\ttt}\Big( \tbbu - \frac{\beta}{3}\tilh^2 \tilde{\Delta}\tbbu
- \beta\tilh (\tilde{\nabla}_H \cdot \tbbu) \nabla_H \tilh \Big)
+ \scrO(\beta^2),
\end{equation*}
which suggests defining 
\begin{equation}
\label{K_GN_3d}
\tilde{\mathbb{K}} = \tilde{\mathbf{\uub}} 
- \frac{\beta}{3 \tilde{h}} \tilde{\nabla}_H 
 \left( \tilde{h}^{3}\mathrm{\tilde{div}}_{H}\tilde{\uub}\right),
\end{equation}
With this definition, the second term can be written as
\begin{align*}
\widetilde{\mathrm{curl}} \big( \tuu_H + \beta \tilde{\nabla}_H \tilh \big) \times \tuu_H
& = 
\left[
\begin{array}{c}
\ddd{\ty} (\tu + \beta \tw \tilh_{\tx})\tv - \ddd{\tx} (\tv + \beta \tw \tilh_{\ty}) \tv \\
\ddd{\tx} (\tv + \beta \tw \tilh_{\ty})\tu - \ddd{\ty} (\tu + \beta \tw \tilh_{\tx}) \tu 
\end{array}
\right]
\\
& = 
\left[
\begin{array}{c}
\Big( \ddd{\ty}\tilde{\mathbb{K}}_1 - \ddd{\tx}\tilde{\mathbb{K}}_2 \Big) \tub \\
\Big( \ddd{\tx}\tilde{\mathbb{K}}_2 - \ddd{\ty}\tilde{\mathbb{K}}_1 \Big) \tvb 
\end{array}
\right]
+
\beta
\frac{\tilh^2}{3}
\left[
\begin{array}{c}
\Big(\frac{\partial \tvb}{\partial \tx}-\frac{\partial \tub}{\partial \ty}
\Big) \tilde{\Delta}_H \tvb\\
\Big(\frac{\partial \tub}{\partial \ty}
                         -\frac{\partial \tvb}{\partial \tx}\Big) \tilde{\Delta}_H \tub
\end{array}
\right] +  \scrO(\beta^2)
\\
& = 
\left[
\begin{array}{c}
\Big( \ddd{\ty}\tilde{\mathbb{K}}_1 - \ddd{\tx}\tilde{\mathbb{K}}_2 \Big) \tub \\
\Big( \ddd{\tx}\tilde{\mathbb{K}}_2 - \ddd{\ty}\tilde{\mathbb{K}}_1 \Big) \tvb 
\end{array}
\right]
+  \scrO(\beta^2),
\qquad \qquad \qquad
\qquad \qquad \qquad
\qquad \quad
\end{align*}
where we have used the fact (shown in the appendix) that
$\frac{\partial \tub}{\partial \ty}
                         -\frac{\partial \tvb}{\partial \tx} =  \scrO(\beta)$.
Finally, the last term in \eqref{equation_nonD} can be written as
\begin{multline*}
\tilde{\nabla}_H \bigg\{
\Big( \tilde{\uub} - \Sfrac{\beta}{3}\tilde{h}^{2}
\Delta\mathbf{\tilde{\uub}} - \beta\tilde{h} (\nabla_H \cdot \mathbf{\tilde{\uub}})
\nabla_H \tilde{h}
 + \scrO(\beta^2) \Big) 
\cdot 
\Big(\tbbu - \beta \Sfrac{\tilh^2}{3} \tilde{\Delta}_H \tbbu + \scrO(\beta^2) \Big)
+ \tilh 
\\
- \ohf \Big( |\tbbu|^2 - 2 \beta \Sfrac{\tilh^2}{3} \tbbu \cdot \tilde{\Delta}\tbbu
                       + \beta |\tilh (\tilde{\nabla}_H\cdot \tbbu) |^2 
+ \scrO(\beta^2) \Big)
\bigg\}
\\
=
\tilde{\nabla}_H \bigg\{
\tilde{\mathbb{K}} \cdot \tbbu + \tilh 
- \ohf \Big( |\tbbu|^2 + \beta |\tilh (\tilde{\nabla}_H\cdot \tbbu) |^2 
\Big)
\bigg\} + \scrO(\beta^2).
\end{multline*}
Thus with the definition of $\mathbb{K}$ as in  \eqref{K_GN_3d},
the relation \eqref{equation_nonD} can be written as
\begin{equation}
\ddd{\ttt} \tilde{\mathbb{K}}
+ \widetilde{\mathrm{curl}}(\tilde{\mathbb{K}}) \times \tbbu
+ \tilde{\nabla}_H \left\{ \tilde{\mathbb{K}} \cdot \tbbu + \tilh  
                            - \ohf \beta \tilh^2 |\tilde{\nabla}_H \cdot \tbbu|^2
\right\} = \scrO(\beta^2).
\label{balanceGN}
\end{equation}
This equation is the approximate form of \eqref{equation_new} which
is valid to the same order of approximation as the Green-Naghdi system.
However, as in the two-dimensional case, it appears that the equation
is actually an exact consequence of the Green-Naghdi equations,
and the $\scrO(\beta^2)$ term in \eqref{balanceGN} may be replaced by zero. 
Then, using the equation \eqref{GN_3d_1_dim}, the balance law \eqref{balanceGN}
may be written in dimensional variables as
\begin{equation}
\label{circ3D}
\frac{\partial \mathbb{K}}{\partial t}
+ \mathrm{curl}(\mathbb{K}) \times \uub
+ \nabla_{H}\Big\{ \mathbb{K} \cdot \uub 
+ gh - \frac{1}{2}\Big( \vert \uub \vert^2
+ \big( \Sfrac{D_H h}{Dt} \big)^2\Big) \Big\} =0,
\end{equation}%
where 
$\mathbb{K} = \uub - \frac{1}{3h}\nabla_H \left( h^3 \nabla_H\cdot \uub \right)$,
is the dimensional form of \eqref{K_GN_3d}.

%
\section{Discussion}
\label{section4}
%
Solutions of the water wave problem satisfy
the exact balance equation
\begin{equation*}
\frac{\partial \big( \mathbf{u}\cdot \frac{\partial \bvphi}{\partial s^i}\big) }{\partial t}
 + \frac{\partial }{\partial s^i}\Big( \frac{p}{\rho}
  + gz - \frac{\vert \uu \vert^2}{2}\Big) = 0,
\end{equation*}
where $s_i$ are Lagrangian coordinates, and $\bvphi$ denotes an ensemble of particle paths.
This equation holds both in two and in three dimensions. In two dimensions, the
above relation gives rise to the differential conservation equation \eqref{newK}
involving the tangential velocity along the free surface which we denoted by $K$.
In three dimensions, the somewhat more complicated equation 
\eqref{equation_new} appears. 

These conservation equations have been studied in the Green-Naghdi approximation
which assumes that surface waves are long when compared to the undisturbed depth
of the fluid, and where terms of order $\beta^2$ or higher in the
long-wave parameter $\beta = h_0 / l$ are disregarded. It was found that
these equations give rise to corresponding approximate balance laws
which nevertheless are exact consequences of the Green-Naghdi equation.
In other words, solutions of the Green-Naghdi system in two or three dimensions
satisfy exactly the conservation laws \eqref{head} and \eqref{circ3D}, respectively.

In this respect, we have completed the physical interpretation of the 
conservation laws connected with the Green-Naghdi equations. To summarize
these in the three-dimensional setting, note that solutions of 
\eqref{GN_3d_1_dim} and \eqref{GN_3d_2_dim}
admit conservation of mass, given by
\begin{equation*}
\frac{\partial h}{\partial t}+\nabla_H \cdot ( h \uub)=0.
\end{equation*}%
Conservation of momentum is represented by
\begin{equation*}
\frac{\partial h \uub} {\partial t}
 + \nabla_H \cdot \left( h \uub \otimes \bar{u} +pI\right) =0,
\end{equation*}%
where
\begin{equation*}
p=\frac{gh^2}{2}+\frac{h^2}{3}\frac{D^2_Hh}{Dt^2}.
\end{equation*}%
Conservation of energy in the three-dimensional Green-Naghdi system
is represented by the following conservation law:
\begin{equation*}
\frac{\partial}{\partial t} \Big\{ h \frac{ \left\vert \mathbf{\bar{u}} \right\vert^2}{2} + E \Big\}
   + \nabla_H  \cdot \Big\{ \uub
                             \Big( h \Sfrac{\vert \mathbf{\bar{u}}\vert^2}{2}
                                     + E + p \Big) \Big\} =0,
\end{equation*}%
where 
\begin{equation*}
E=\frac{gh^2}{2} + \frac{h}{6}\Big( \frac{D_Hh}{Dt}\Big)^2.
\end{equation*}%
Finally, the equations admit the following exact consequence:
\begin{equation*}
\frac{\partial \mathbb{K}}{\partial t}+\mathrm{curl}(\mathbb{K}) \times \mathbf{\bar{u}}
+\nabla_H \Big\{ \mathbb{K}\mathbf{\cdot \bar{u}} 
               + gh - \frac{1}{2}\Big( \vert \mathbf{\bar{u}} \vert ^2 
                    +\big( \Sfrac{D_Hh}{Dt}\big)^ 2 \Big) \Big\} =0
\end{equation*}%
with 
\begin{equation*}
\mathbb{K} = \uub + \frac{1}{3h}\nabla_H \Big( h^{2}\frac{D_Hh}{Dt} \Big) .
\end{equation*}%
This conservation law is obtained as an approximation of a 
general exact balance law \eqref{modified_K} for the full Euler equations.
In \cite{GT} such a balance law was derived and interpreted as the analogue
of the Bernoulli integral even in a more general case covering, in particular, 
applications to multi-phase flow modeling.
The importance of the variable $\mathbb{K}$ for numerical purposes is demonstrated in 
\cite{LGLX,Metayer10}.
Analogous equations can also be obtained in the case of multi-layer flows,
such as for instance in \cite{BGT} and \cite{PCH}. \\

{\bf Acknowledgments.}
SG was partially supported by ANR BoND, France. 
HK and ZK acknowledge support by the Research Council of Norway.

\appendix
\section{}\label{appA}
%
%
\subsection{Derivation of the 2D Green-Naghdi equations}
%
%
%
The derivation of the Green-Naghdi equations in two spatial dimensions
can be found in \cite{Barth04}. Here we just recall some of the key steps.
Using the scaling \eqref{scaling}, and
averaging the continuity equation \eqref{Euler_2}
over the local depth of the fluid leads to the continuity equation 
\eqref{GN_1},
where the average horizontal velocity is defined by
$\tub =\frac{1}{\tilh}\int_{0}^{\tilh} \tu d\tz.$
%
Using the boundary conditions, the continuity equation and depth averaged
values, the momentum equation 
\begin{equation*}
\tub_{\ttt} + \tub \tub_{\tx} + \tilde{\eta}_{\tx} 
          + \frac{\beta }{\tilh} \frac{\partial }{\partial \tx}
\int_{0}^{\tilh} \tz \Gamma (\tx,\tz,\ttt) \, d\tz=
\frac{-1}{\tilh} \frac{\partial }{\partial \tx} \int_{0}^{\tilh} 
\left( \tu^2 - \tub^2\right) \,dz
\end{equation*}
appears, where 
\begin{equation*}
\Gamma (\tx,\tz,\ttt) =\tw_{\ttt} + \tu \tw_{\tx} + \tw\tw_{\tz}
\end{equation*}
is the vertical acceleration.
The velocity potential is expanded as an asymptotic power series in the
vertical coordinate around $\tz = 0$. Then the horizontal
velocity in terms of the average velocity is 
\begin{equation*}
\tu(\tx,\tz,\ttt) = \tub + \frac{1}{6}\beta \tilh^2
                    \frac{\partial ^2 \tub}{\partial \tx^2} 
                   - \frac{1}{2} \beta \tz^2 \frac{\partial ^2 \tub}{\partial \tx^2}
                  + \scrO(\beta ^2).
\end{equation*}
Moreover, it can be shown that 
\begin{equation*}
\int_{0}^{h}\left( \tu^2 - \tub^2\right) \,d\tz = \scrO(\beta ^2),
\end{equation*}
and that the vertical velocity becomes 
\begin{equation*}
\tw(\tx,\tz,\ttt) = -\tz \frac{\partial \tub}{\partial \tx} + \scrO(\beta ).
\end{equation*}
Therefore, 
\begin{equation*}
\Gamma (\tx,\tz,\ttt) = -\tz \left( \tub_{\tx\ttt} 
                        + \tub\tub_{\tx\tx}-\tub_{\tx}^2 \right) + \scrO(\beta ),
\end{equation*}
and we find the equation 
\begin{equation*}
\tub_{\ttt} + \tub\tub_{\tx} + \tilde{\eta}_{\tx} 
          - \frac{\beta }{3\tilh} \frac{\partial }{\partial \tx} \left( \tilh^3 \tub_{\tx\ttt} 
          + \tub\tub_{\tx\tx} - \tub_{\tx}^2\right) =\scrO(\beta ^{2}).
\end{equation*}
By ignoring terms of order $\beta^2$ and rescaling, \eqref{GN_2} is obtained.

%
%
\subsection{Derivation of the 3D Green-Naghdi equations}
In the three-dimensional case we use the additional scalings
\begin{equation*}
\quad \tilde{y}=\frac{y}{l} \  \mbox{  and  }  \ \tilde{v}=\frac{v}{c_0},
\end{equation*}
so that we can write the horizontal velocity in the form
$\tuu_H = \frac{1}{c_0}\uu_H$.
Averaging over the
local depth of the fluid leads to the continuity equation 
\begin{equation}
\label{GN_3d_1}
\tilde{\eta}_{\ttt}+ \tilde{\nabla}_H \cdot( \tilh \tbbu ) = 0,
\end{equation}
where $\tbbu =(\tub,\tvb)$ is the averaged non-dimensional horizontal velocity:
$\bar{\tilde{\mathbf u}} = \tbbu = \frac{1}{\tilh} \int_0^{\tilh} \tuu_H \, d \tz.$
Using the boundary conditions, the
continuity equation and depth averaged values, the momentum equations 
\begin{multline*}
\tub_{\ttt} + ( \tbbu \cdot \tilde{\nabla}_H ) \tub
            + \tilde{\eta}_{\tx} 
  + \frac{\beta}{\tilh} \frac{\partial}{\partial \tx}\int^{\tilh}_{0}\tz \Gamma(\tx,\ty,\tz,\ttt)\,d \tz = \\
   \frac{-1}{\tilh} \frac{\partial}{\partial \tx} \int^{\tilh}_0 \left(\tu^2 - \tub^2 \right)\,d\tz 
    - \frac{1}{\tilh}\frac{\partial}{\partial \ty} \int^{\tilh}_0 \left( (\tu\tv)^2
    - \tub^2 \right)\,d\tz,
\end{multline*}
and 
\begin{multline*}
\tilde{\bar{v}}_{\ttt} + ( \tbbu \cdot \tilde{\nabla}_H ) \tv 
                     +  \tilde{\eta}_{\ty} 
  + \frac{\beta}{h} \frac{\partial}{\partial \tx} \int^{\tilh}_0 \tz \Gamma(\tx,\ty,\tz,\ttt)) \, d \tz = \\
\frac{-1}{\tilh}\frac{\partial}{\partial \tx} \int^{\tilh}_0 \left(\tv^2 - \tilde{\bar{v}}^2 \right) \, d\tz 
     -\frac{1}{\tilh}\frac{\partial}{\partial \ty} \int^{\tilh}_0 \left((\tu\tv)^2
      -(\tilde{\bar{v}})^2 \right) \, d\tz,
\end{multline*}
appear, where 
\begin{equation*}
\Gamma(\tx,\ty,\tz,\ttt) = \tw_{\ttt} + \tu\tw_{\tx}  + \tv\tw_{\ty} + \tw\tw_{\tz}.
\end{equation*}

The velocity potential is expanded as an asymptotic power series in the
vertical coordinate around the bottom $\tz=0$. Then the horizontal
velocity in terms of the average velocity is 
\begin{equation*}
\label{HorizontalVelocityField}
\tilde{\mathbf{u}} (\tx,\ty,\tz,\ttt) = \tbbu 
           + \beta \Big( \frac{1}{6} h^{2}-\frac{1}{2} \tilde{z}^{2} \Big) %
\Delta_H \tbbu + \scrO(\beta^{2}).
\end{equation*}
Moreover, it can be shown that 
\begin{equation*}
\int^{\tilh}_0 \left(\tu^2 - \bar{\tilde{u}}^2 \right)\,d\tz = \scrO(\beta^2), 
\quad \int^{\tilh}_0 \left(\tv^2- \bar{\tilde{v}})^2 \right) \, d\tz = \scrO(\beta^2), 
\quad \int^{\tilh}_0 \left((\tv\tu)^2 - (\bar{\tilde{u}})^2 \right) \,d \tz=\scrO(\beta^2),
\end{equation*}
and that the vertical velocity becomes 
\begin{equation*}
\tv(\tx,\tz,\ttt)= -\tz \tilde{\nabla}_H \cdot \tilde{\bar{\mathbf{u}}} + \scrO(\beta).
\end{equation*}
Therefore, 
\begin{equation*}
 \Gamma(\tx,\ty,\tz,\ttt) = - \tz\left[(\bar{\tilde{u}}_{\tilde{x}}
                           + \bar{\tv}_{\ty})_{\ttt} + \bar{\tilde{u}} \bar{\tilde{u}}_{\tilde{x}\tilde{x}}
                           + \bar{\tu}\bar{\tv}_{\ty\tx} 
                           + \bar{\tv}\bar{\tu}_{\tx\ty} 
                           + \bar{\tv}\bar{\tv}_{\ty\ty} 
                           - (\bar{\tilde{u}}_{\tx}
                           + \bar{\tilde{u}}_{\ty}^2 \right] +\scrO(\beta),
\end{equation*}
and we find the equation 
\begin{equation}
\big\{ 1+\beta \tilde{\mathcal{T}}(\tilh, \tbbu) \big\} \tbbu_{\tilde{t}} 
+ \tbbu \cdot \tilde{\nabla}_H \tbbu + \tilde{\nabla}_H \tilde{\eta} 
+ \beta Q(h, \tbbu) = \scrO(\beta^2)
\label{GN_3d_2},
\end{equation}
where 
\begin{equation*}
\tilde{\mathcal{T}}(h,V) = -\frac{1}{3h} \tilde{\nabla}_H(h^3 \tilde{\nabla}_H \cdot V),
\end{equation*}
\begin{equation*}
\tilde{Q}(h,V)=\frac{-1}{3h}\tilde{\nabla}_H \Big[h^3\Big( V \cdot \tilde{\nabla}_H (\tilde{\nabla}_H \cdot V )
 -( \tilde{\nabla}_H \cdot V)^2\Big)\Big].
\end{equation*}

\subsection{Alternative representation of the horizontal velocity}
An alternative representation of the horizontal velocity field can be obtained
by using the relation
\begin{equation*}
\tilde{\Delta}_{H}\tilde{\bar{\mathbf{u}}}= \tilde{\nabla}_{H}\left( \nabla _{H}\cdot 
\tilde{\bar{\mathbf{u}}}\right) +\scrO(\beta ),
\end{equation*}%
which holds for potential flows.
Indeed, note that
\begin{eqnarray*}
\tilde{\nabla}_{H}\left( \tilde{\nabla}_{H} \cdot \tbbu \right) 
& = & \left( \frac{\partial }{\partial \tx}\left( \frac{\partial \tub}{\partial \tx} 
+ \frac{\partial \tilde{\bar{v}}}{\partial \ty}\right) 
,
\frac{\partial }{\partial \ty}\left( 
\frac{\partial \tub}{\partial \tx}+\frac{\partial \tilde{\bar{v}}}{\partial \ty}\right) \right)
\\
& = & \left( \frac{\partial^2 \tub}{\partial \tx^2} +\frac{\partial^2\tub}{\partial \ty^2}
-\frac{\partial }{\partial \ty}\left( \frac{\partial \tub}{\partial \ty}
-\frac{\partial \tilde{\bar{v}}}{\partial \tx}\right) 
,
\frac{\partial^2 \tilde{\bar{v}}}{\partial \tx^2} + \frac{\partial^2 \tilde{\bar{v}}}{\partial \ty^2}
+\frac{\partial }{\partial \tx}\left( \frac{\partial \tub}{\partial \ty}
-\frac{\partial \tilde{\bar{v}}}{\partial \tx}\right) \right)
\\
& = & \tilde{\Delta}_{H} \tbbu +
\left( \frac{\partial \tilde{\bar{\omega}}}{\partial y}
,
- \frac{\partial \tilde{\bar{\omega}}}{\partial x}\right),
\end{eqnarray*}%
where we have introduced the the vorticity of mean horizontal velocity:
\begin{equation*}
\tilde{\bar{\omega}} 
= \frac{\partial \tilde{\bar{v}}}{\partial \tx}
- \frac{\partial \tub}{\partial \ty}.
\end{equation*}
Let us estimate $\tilde{\bar{\omega}}$ under the condition that the instantaneous flow
is irrotational: $v_{x}-u_{y}=0$. We have 
\begin{eqnarray*}
\tilde{\bar{\omega}} 
& = & \frac{\partial }{\partial \tx}\left( \frac{1}{\tilh} \int_{0}^{\tilh} \tv d\tz \right)
- \frac{\partial }{\partial \ty}\left( \frac{1}{\tilh} \int_{0}^{\tilh} \tu d\tz \right) 
\\
& = & \frac{\tilh_{\tx}}{\tilh} \tilde{\bar{v}} - \frac{\tilh_{\ty}}{\tilh}\tub 
+  \frac{1}{\tilh} \int_{0}^{\tilh} \left( \tv_{\tx} - \tu_{\ty}\right) d\tz
+ \frac{\tilh_{\tx}}{\tilh}\tv \left(\ttt,\tx,\ty,\tilh(\ttt,\tx,\ty)\right)
- \frac{\tilh_{\ty}}{\tilh} \tu\left(\ttt,\tx,\ty,\tilh(\ttt,\tx,\ty) \right) 
\\
& = & - \frac{\tilh_{\tx}}{\tilh}\left( \tilde{\bar{v}}- \tv \left(\ttt,\tx,\ty,\tilh(\ttt,\tx,\ty)\right) \right) 
+ \frac{\tilh_{\ty}}{\tilh}\left( \tub - \tu \left(\ttt,\tx,\ty,\tilh(\ttt,\tx,\ty)\right) \right).
\end{eqnarray*}%
Now from \eqref{HorizontalVelocityField}, we see that
\begin{equation*}
\left\vert \tilde{\bar{u}}-\tilde{u}\left( t,x,y,h(t,x,y)\right) \right\vert
=O\left( \beta \right),
\end{equation*}%
and
\begin{equation*}
\left\vert \tilde{\bar{v}}-\tilde{v}\left( t,x,y,h(t,x,y)\right) \right\vert
=O\left( \beta \right).
\end{equation*}%
Hence, the vorticity of mean horizontal velocity can be seen to be of the order of $\beta$.
As a consequence, one can also use the following approximate formula for the
horizontal velocity field at the free surface in terms 
of the average horizontal velocity:
\begin{equation*}
\tuu_{H} ( \tilde{\mathbf{x}},\tilh,\ttt) 
= \tilde{\bar{\mathbf{u}}}
- \beta \frac{\tilh^2}{3} \tilde{\nabla}_{H} \left( \tilde{\nabla}_{H} \cdot \tbbu \right) 
+ \scrO(\beta^2).
\end{equation*}%

\end{document}